\documentclass[reprint,prl,letterpaper,superscriptaddress]{revtex4-1}
\usepackage[applemac]{inputenc}
\usepackage{graphicx}
\usepackage{dcolumn}
\usepackage{bm}
\usepackage{amsmath,amssymb}
\usepackage{pifont} 
\usepackage{graphicx}

\usepackage{epstopdf}
\usepackage{bm}
\usepackage[normalem]{ulem} 
\usepackage{gensymb} 
\usepackage[usenames,dvipsnames]{color}

\definecolor{OliColor}{rgb}{0.1,0.1,0.8} 
\definecolor{OliComColor}{rgb}{0.5,0.5,1} 

\definecolor{AbeColor}{rgb}{0.8,0.1,0.1} 
\definecolor{AbeComColor}{rgb}{1,0.5,0.5} 

\definecolor{SupMatColor}{rgb}{0.85,.66,0}

\begin{document}

\graphicspath{}

\title{Capillarity-Driven Flows at the Continuum Limit}

\author{Olivier Vincent}
\email{orv3@cornell.edu}
\affiliation{School of Chemical and Biomolecular Engineering, Cornell University, Ithaca, New York 14853, USA}

\author{Alexandre Szenicer}
\affiliation{School of Chemical and Biomolecular Engineering, Cornell University, Ithaca, New York 14853, USA}

\author{Abraham D. Stroock}
\email{ads10@cornell.edu}
\affiliation{School of Chemical and Biomolecular Engineering, Cornell University, Ithaca, New York 14853, USA}
\affiliation{Kavli Institute at Cornell for Nanoscale Science, Cornell University, Ithaca, New York 14853, USA}


\begin{abstract}
We experimentally investigate the dynamics of capillary-driven flows at the nanoscale, using an original platform that combines nanoscale pores and microfluidic features. Our results show a coherent picture across multiple experiments including imbibition, poroelastic transient flows, and a drying-based method that we introduce. In particular, we exploit extreme drying stresses -- up to 100 MPa of tension -- to drive nanoflows and provide quantitative tests of continuum theories of fluid mechanics and thermodynamics (e.g. Kelvin-Laplace equation) across an unprecedented range. We isolate the breakdown of continuum as a negative slip length of molecular dimension. 
\end{abstract}

\maketitle




Fluids confined at the nanoscale play central roles in many areas of science and technology, from geophysics \cite{Saadatpoor2010,Birdsell2015},
plant hydraulics and biomaterials \cite{Stroock2014,Fornasiero2008} to catalysis and filtration \cite{Li2010,Mohmood2013}. These contexts have motivated a considerable effort to understand both the dynamic \cite{Huber2015,Bocquet2010}  and thermodynamic \cite{Gelb1999,Findenegg2008} behavior of nano-confined liquids, but fundamental points of debate persist. With respect to dynamics, theoretical considerations suggest that continuum fluid mechanics (Navier-Stokes) should govern flows in conduits greater than $\sim 1$ nm in lateral dimension \cite{Bocquet2010} but significant uncertainty remains with respect to both constitutive properties and boundary conditions. For example: experimental measurements of the shear viscosity of water have only constrained it to within a factor of three for confinement below $4$ nm \cite{Raviv2001} and, while simulations indicate that slip lengths should tend to zero on wetting surfaces \cite{Huang2008} experimental values typically span $-5$ to $5$ nm \cite{Bocquet2010,Honig2007},
leaving open the possibility of strongly affected flows in nanometric pores. 
With respect to thermodynamics, uncertainty remains regarding the limits of validity of the macroscopic picture of liquid-vapor phase behavior (Kelvin-Laplace) in pores below 10 nm-diameter \cite{Neimark2003}
with increasing importance of solid-fluid interactions (e.g., disjoining pressure) and the emergence of inhomogeneities in the form of molecular layering and anisotropic stresses \cite{Long2013}. However excellent agreement have also be found between macroscopic predictions and simulation results for desorption in pores $<3$ nm in diameter \cite{Factorovich2014_2}.
In a situation that couples transport and thermodynamics, capillary-driven flow rates measured by numerous groups deviate significantly (up to 2 fold) from predictions based of macroscopic theory for nanochannels of height below ~100 nm \cite{Chauvet2012}.

These large uncertainties result in part from the challenges associated with the study of highly confined liquids, such as the difficulty of performing direct measurements of the liquid state at the pore-scale and of generating measurable flows given the small volumes involved and the extreme viscous drag \cite{Bocquet2010}. 
Surface tension driven flows provide an attractive opportunity due to the large capillary driving forces that can develop spontaneously in nanopores. Imbibition, for example, is a common process where a wetting liquid spontaneously invades a porous medium,
and has received recent attention as a tool in nanofluidics \cite{GruenerPRL2009,Acquaroli2011}. Imbibition, however, is an unsteady process and offers little control over the flow and thermodynamic state of the liquid. 
Drying is another capillary-related process involved in many situations from the circulation of sap in plants \cite{Stroock2014} to biomaterials in contexts such as wound dressings or contact lenses \cite{Fornasiero2008}. However, despite considerable work on the macroscopic fluid mechanics of drying \cite{Chauvet2009,LehmannOr2008} on the one hand, and on the thermodynamics of desorption from nanopores on the other hand \cite{Ravikovitch2002,Wallacher2004}, the dynamics of drying at the nanoscale has only received recent attention \cite{Duan2012,Vincent2014} and remains largely unexplored.

In this Letter, we investigate drying as an original way to generate controlled flows of highly confined liquids. We propose an experimental platform that couples nanofluidic and microfluidic structures to allow for precise measurements of both drying and imbibition with nanoscale confinement. We obtain a coherent picture of capillarity-induced flows at the nanometer scale across experimental methods and various liquids. In particular, we show quantitatively that the tension (negative pressure) in the liquid predicted by the thermodynamics of unsaturated liquid-vapor equilibria (Kelvin equation) manifests itself as a hydraulic stress out to $\simeq 100$ MPa. This tensile stress drives macroscopic flows through a material that presents massive hydraulic resistance, suggesting the power of drying-based techniques to study liquids at the nanoscale. We also show that the resulting permeation follows the continuum laws of capillary-viscous flows provided that a simple correction is included to account for molecular adsorption ("sticking") at the pore wall. Our results thus provide an unusually robust experimental test of transport and thermodynamic models at the continuum limit.







\begin{figure}[]
	\begin{center}
	\includegraphics{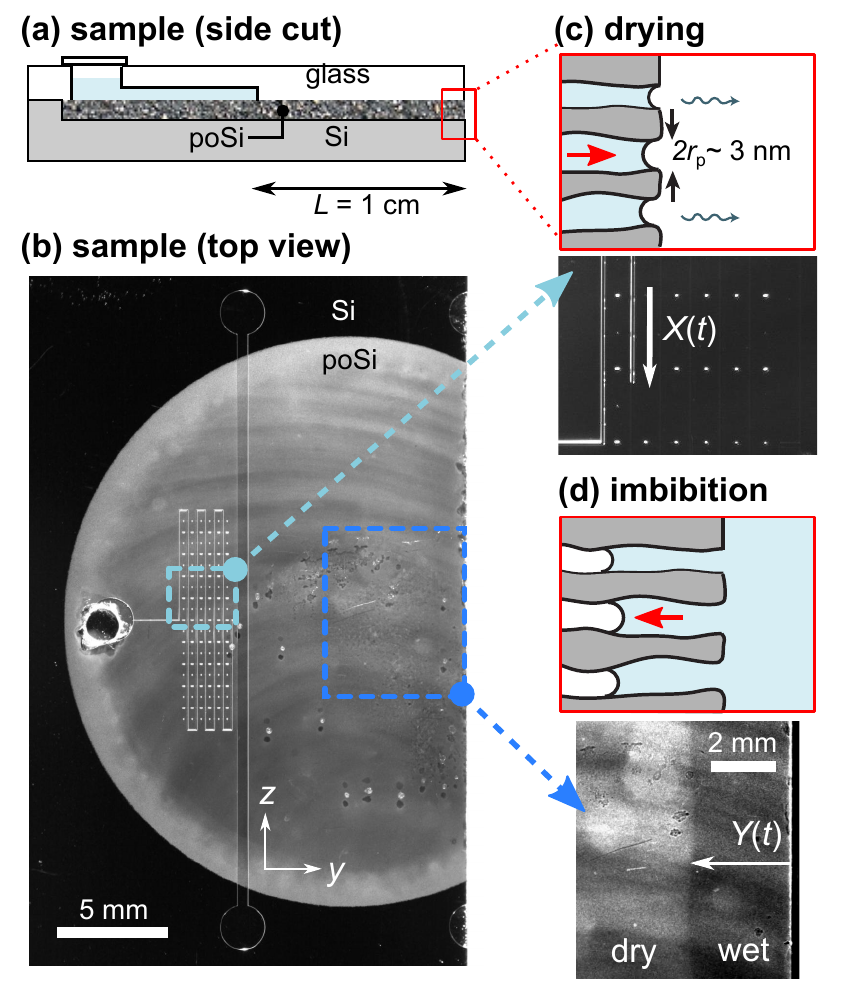}
			\caption{\small (a) Side-view sketch and (b) top view photograph of the micro/nano-fluidic platform. (b) The drying flux was measured by the position, $X(t)$ of the air/water interface in the serpentine-shaped reservoir, while (b) imbibition dynamics was measured optically from the position of the wetting front, $Y(t)$ that was visible optically.
 }
\label{fig:SampleSetup}
\end{center}
\end{figure}


Figure \ref{fig:SampleSetup} presents our sample. Details of the fabrication and of the experimental procedure can be found in the Supplemental Material \footnote{See Supplemental Material available at http://www.stroockgroup.org/home/publications}.
Briefly, we used anodization to etch a circular zone in a silicon wafer, leading to a $5.5\,\mathrm{\mu}$m-thick layer of isotropic, interconnected pores with a distribution of pore sizes in the range, $r_\mathrm{p}=1.4 \pm 0.4$ nm and porosity $\phi=0.45$ based on $\mathrm{N_2}$ porosimetry. We used lithography to form a serpentine-shaped reservoir in a glass wafer and drilled a hole to access the reservoir through the glass. Finally we bonded silicon to glass and opened the pores to the outside at the edge of the sample by dicing (Fig \ref{fig:SampleSetup}a-b). 
We performed drying experiments (Fig \ref{fig:SampleSetup}c) by filling the pores and reservoir with liquid, and then placing the sample in a vapor of controlled humidity after sealing the inlet hole. By tracking the air-water meniscus position in the serpentine reservoir ($X(t)$ in Fig \ref{fig:SampleSetup}c) we could evaluate the flow rate through the porous layer due to evaporation from the open edge.
For imbibition experiments, we placed evacuated samples in the liquid of interest and monitored the advancing wetting front ($Y(t)$ in Fig \ref{fig:SampleSetup}d) by time-lapse photography. 

In order to describe the thermodynamic states of liquid and vapor in a unified way, we use  \emph{liquid potential}, $\Psi=(\mu-\mu_0)/v_\mathrm{m}$ \cite{Stroock2014} where $\mu \mathrm{[J / mol]}$ is the chemical potential and $v_\mathrm{m}(P,T)\,\mathrm{[m^3 / mol]}$ is the liquid molar volume. This quantity expresses the chemical potential of each phase as the corresponding pressure of the pure, bulk liquid. Neglecting the pressure dependence of $v_\mathrm{m}$, an approximation that we expect to hold to within $2.6  \%$ for water \cite{Note1},
\begin{equation}
  \begin{cases}
   \Psi_\mathrm{liq} = P-P_0 & \text{for the liquid} \\
   \Psi_\mathrm{vap} = \frac{RT}{v_\mathrm{m}}\ln\left(\frac{p}{p_\mathrm{sat}}\right) & \text{for the vapor}
  \end{cases}
\label{eq:WaterPotential}
\end{equation}
where $P$ and $p$ represent respectively the liquid and the partial vapor pressures, $P_0$ is atmospheric pressure, $p_\mathrm{sat}(T,P_0)$ is the saturated vapor pressure and $RT$ is the thermal energy.

\begin{figure}[]
	\begin{center}
	\includegraphics{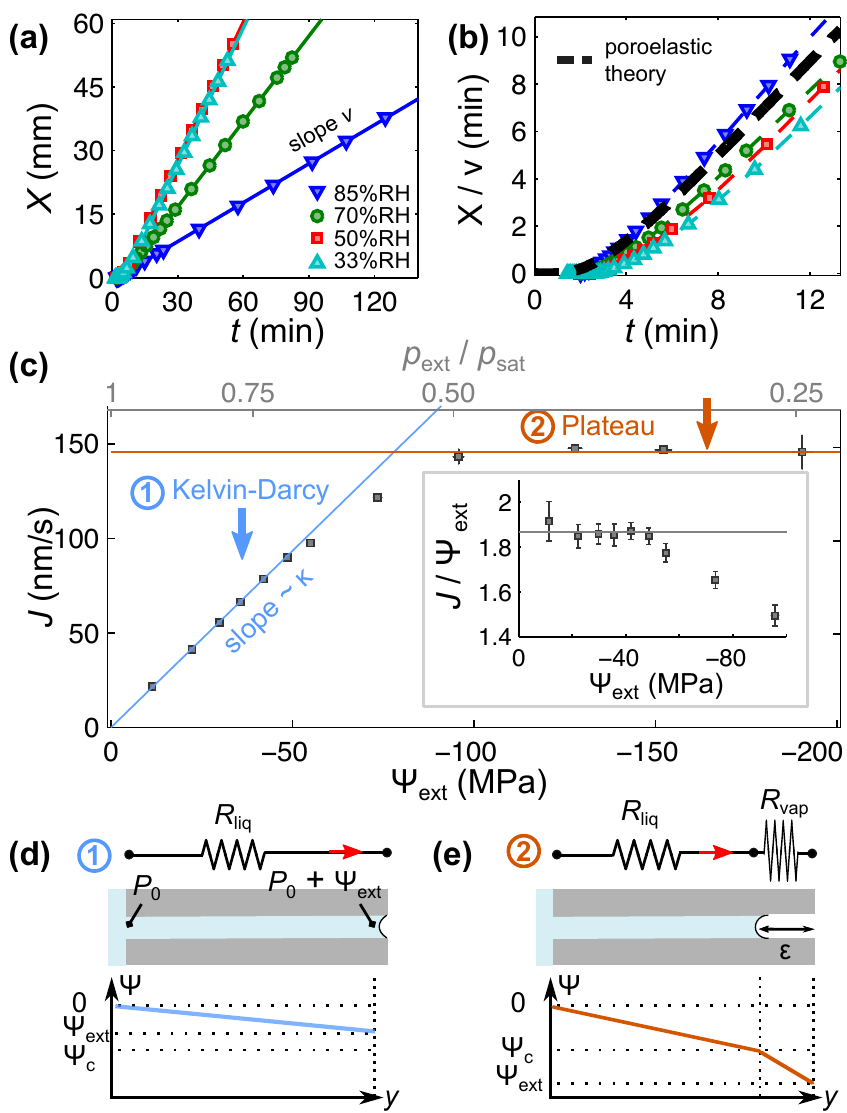}
			\caption{\small Drying dynamics with water. (a) Raw data for the meniscus position, $X(t)$ (Fig. \ref{fig:SampleSetup}c). (b) Data in (a) normalized by the steady-state speed $v$ to show the transients. The dashed, black curve shows the prediction from the theory of poroelasticity \cite{Note1}. (c) Volumetric flux (Darcy velocity) as a function of the external vapor potential, $\Psi_\mathrm{ext}$ (bottom axis) calculated with Eq. (\ref{eq:Kelvin}) from the imposed relative humidity $p_\mathrm{ext}/p_\mathrm{sat}$ (top axis). The lines represent the two predicted regimes from Eqs. (\ref{eq:Darcy}) and  (\ref{eq:FluxPlateau}). (d-e) Schematic diagrams of our model for these two regimes, with sketches of the spatial profiles of the liquid potential.
 }
\label{fig:Drying}
\end{center}
\end{figure}

We first recorded drying dynamics with the position of the meniscus $X(t)$ at a variety of humidities (Fig \ref{fig:Drying}a). The results show the establishment of a steady-state after a transient of a few minutes (Fig \ref{fig:Drying}b). From the speed, we used the geometry of the porous layer to calculate the steady-state volumetric flux in the porous medium (Darcy velocity), $J \, \mathrm{[m/s]}$. Data as in Fig. \ref{fig:Drying}c revealed two regimes as a function of liquid potential.

We begin by considering the linear regime (\ding{172} in Fig. \ref{fig:Drying}c) with the physical picture of the pore-scale depicted in Fig. \ref{fig:Drying}d. Here, we assume that the menisci remained pinned at the open edge of the porous layer and that local equilibrium was established between the pore liquid behind the mensci and the bulk vapor (partial pressure $p_\mathrm{ext})$; we expect this equilibrium to hold to good approximation due to the large transport resistance in the porous medium relative to that in the vapor \cite{Note1}.  To evaluate this equilibrium, we suppose that, despite the strong confinement imposed by the pores, the properties of the liquid
are unperturbed relative to that of a bulk phase.  In other words, the pore liquid responds to the lowering of the vapor pressure with a reduction of its hydrostatic pressure, as predicted by the well-known Kelvin equation ($\Psi_\mathrm{liq} = \Psi_\mathrm{vap}$ from Eq. \ref{eq:WaterPotential}).  
\begin{equation}
P_\mathrm{edge}-P_0=\Psi_\mathrm{ext} = \frac{RT}{v_\mathrm{m}}\ln\left(\frac{p_\mathrm{ext}}{p_\mathrm{sat}}\right).
\label{eq:Kelvin}
\end{equation}
Eq. (\ref{eq:Kelvin}) provides an expression for the pressure difference that drives flow through the layer of porous silicon between the microchannel and the open edge and shows that this driving force is directly related to the vapor humidity through thermodynamics.  Here, we neglect both the pressure drop along the microchannel and the capillary pressure associated with the macroscopic meniscus at the channel's left end, such that the pressure of liquid entering the porous layer is approximately atmospheric ($P_0$) \cite{Note1}. Assuming Darcy flow \cite{Bear1988} through the porous medium, we thus predict linear response of the permeation rate with respect to $\Psi_\mathrm{ext}$:
\begin{equation}
J =-\kappa(P_\mathrm{edge}-P_0) / L=-\kappa \Psi_\mathrm{ext} / L.
\label{eq:Darcy}
\end{equation}
where $\kappa$ is the Darcy permeability of the porous layer and $L$ is the geometrical length of the permeation path (Fig. \ref{fig:SampleSetup}). The blue line and inset in Fig. \ref{fig:Drying}c clearly illustrate the robustness of this linear regime down to $\Psi_\mathrm{ext} \simeq -50$ MPa.  We extract the permeability from the best fit slope: $\kappa=(1.87 \pm 0.08) \times 10^{-17} \, \mathrm{m^2/(Pa.s)}$, a value that is in the lower range of permeability of rocks considered impervious such as granite \cite{Bear1988}.
We compare this result to Carman-Kozeny theory
\begin{equation}
\kappa=\frac{\phi \, r_\mathrm{eff}^2}{8 \, \eta \, \tau}
\label{eq:Carman}
\end{equation}
where $r_\mathrm{eff}$ is an effective pore size \cite{Gruener2009}. The tortuosity $\tau$, although often used as an adjustable factor to fit Eq. (\ref{eq:Carman}) to experimental results,
is expected to take physically reasonable values close to $\tau=4$ for isotropic pore architectures \cite{Gruener2009}. Using the value of $\kappa$ experimentally determined above and $\tau=4$, we find a preliminary estimate of $r_\mathrm{eff} \simeq 1.1$ nm, which lies in the lower range of the pore sizes estimated from $\mathrm{N_2}$ porosimetry. 



Independently of the pore geometry, we can use the transient responses (Fig. \ref{fig:Drying}b) to perform a second check of the permeability and of deviations of the properties of the pore liquid from those of the bulk. Poroelastic theory predicts that pressure diffuses through a porous medium with a diffusivity $C=\kappa/(\phi \chi)$ where $\chi$ is the effective compressibility of the fluid-filled solid. To generate the black curve in Fig. 2b, we used $\kappa$ estimated above and took for $\chi$ the bulk compressibility of liquid water ($4.5\times10^{-10}\,\mathrm{Pa}^{-1}$) \cite{Note1}; the agreement with experiments indicates that we have measured a coherent value of the permeability and that confinement has not strongly perturbed the thermodynamic properties of the pore liquid.



We now consider the plateau regime of the flow at larger driving forces (\ding{173} in Fig. \ref{fig:Drying}c) with the physical picture depicted in Fig. \ref{fig:Drying}e.
Once the pressure difference required for equilibrium with the bulk vapor (Eq. \ref{eq:Kelvin}) exceeds the capillary pressure that can be maintained by the pores, we suppose that the menisci recede from the edge and that the flow within the porous medium becomes multiphase -- liquid between $y=0$ and $L - \epsilon$ (linear mass conductance $g_\mathrm{liq}$ [s], see \cite{Note1}), vapor between $y=L - \epsilon$ and $L$ ($g_\mathrm{vap}$). Using the Laplace equation for the capillary pressure, this retreat of the menisci occurs when
\begin{equation}
\Psi_\mathrm{c}=-2\sigma\cos\theta/r_\mathrm{p}.
\label{eq:Laplace}
\end{equation}
For conditions drier than $\Psi_\mathrm{c}$, the flow thus experiences resistors-in-series in response to an imposed driving force $\Psi_\mathrm{ext}$, with variable resistors of values $R_\mathrm{liq}=(L-\epsilon)/g_\mathrm{liq}$ and $R_\mathrm{vap}=\epsilon / g_\mathrm{vap}$. The variable position of the meniscus $\epsilon$ is found by satisfying $\Psi(L-\epsilon)= \Psi_c$.
Based on Darcy's law for the liquid phase and Knudsen diffusion for the vapor phase, we estimate $g_\mathrm{vap}/g_\mathrm{liq}\simeq10^{-4}$ which implies $\epsilon / L \sim g_\mathrm{vap}/g_\mathrm{liq} \ll 1$  \cite{Note1}. It follows that the system reaches a steady-state with the vapor-liquid interface close to the edge and with a pressure gradient in the liquid phase equal to $\Psi_\mathrm{c}/(L-\epsilon) \simeq \Psi_\mathrm{c}/L$, leading to
\begin{equation}
J_\mathrm{c} = -\kappa \Psi_\mathrm{c} / L
\label{eq:FluxPlateau}
\end{equation}
to excellent approximation. The flow predicted by Eq. (\ref{eq:FluxPlateau}) is independent of the imposed external humidity; this independence explains the plateauing of the drying flux observed experimentally (Fig. \ref{fig:Drying}c).
In this regime the driving force for the flow is set by the intrinsic capillary pressure of the pores $\Psi_\mathrm{c}$ instead of Kelvin pressure $\Psi_\mathrm{ext}$.

Comparing Eqs. (\ref{eq:Darcy}) and (\ref{eq:FluxPlateau}), we remark that $\Psi_\mathrm{c}$ corresponds to the intersection point between the two regimes \ding{172} and  \ding{173} and thus provides an independent measure of the capillary pressure $\Psi_\mathrm{c}=-76\pm5$ MPa. Using Laplace law (Eq. \ref{eq:Laplace}) with $\theta=25\pm5^\circ$ \cite{Note1} then yields a new pore size estimate $r_\mathrm{p,c}=1.7\pm0.2$ nm. This value is in very good agreement with the range estimated from $\mathrm{N_2}$ porosimetry, but there is an apparent inconsistency with the pore radius $r_\mathrm{eff}=1.1$ nm deduced from Carman-Kozeny equation (\ref{eq:Carman}).



\begin{figure}[]
	\begin{center}
	\includegraphics{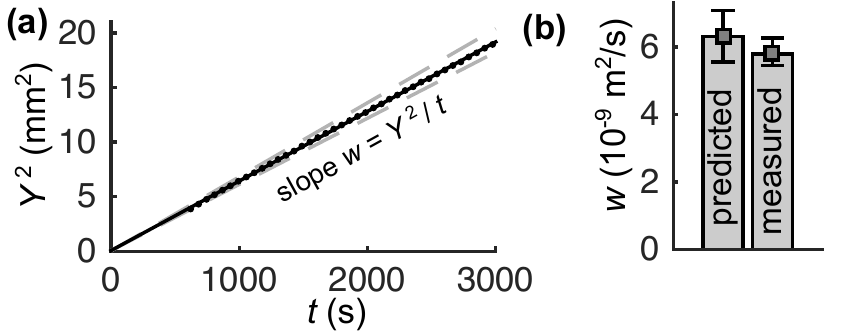}
			\caption{\small Imbibition with water. (a) Front position $Y^2(t)$. The grey dashed lines represent the uncertainty from the image analysis \cite{Note1}. (b) Comparison between the measured imbibition speed $w$ and the  prediction from Eq. (\ref{eq:Washburn}), using the values of $\kappa$ and $\Psi_\mathrm{c}$ deduced from the drying experiments.
			 }
\label{fig:Imbibition}
\end{center}
\end{figure}


In order to resolve this conflict and check the validity of our interpretation of the drying regimes, we measured imbibition dynamics in the same sample. 
During imbibition of a dry sample, the liquid front position is expected to follow $Y(t)=\sqrt{w \, t}$ with
\begin{equation}
w=-\frac{2\,\kappa\, \Psi_\mathrm{c}}{\phi}
\label{eq:Washburn}
\end{equation}
known as the Lucas-Washburn law \cite{WickingBook2013}. Interestingly, the two important parameters governing imbibition dynamics are precisely the $\kappa$ and $\Psi_\mathrm{c}$ estimated from the drying analysis above. This correspondence emphasizes the similar physics involved in both drying and imbibition: a combination of capillarity and viscous flow at the nanoscale.
Fig. \ref{fig:Imbibition}a shows that imbibition in our system followed the $\sqrt{t}$ scaling and agreed with Eq. (\ref{eq:Washburn}), using the values of $\kappa$ and $\Psi_\mathrm{c}$ estimated from drying (Fig. \ref{fig:Imbibition}c). This result confirms the appropriateness of our interpretation of the drying regimes and suggests that the inconsistencies in the pore size estimates described above have a more fundamental source.



\begin{figure}[]
	\begin{center}
	\includegraphics{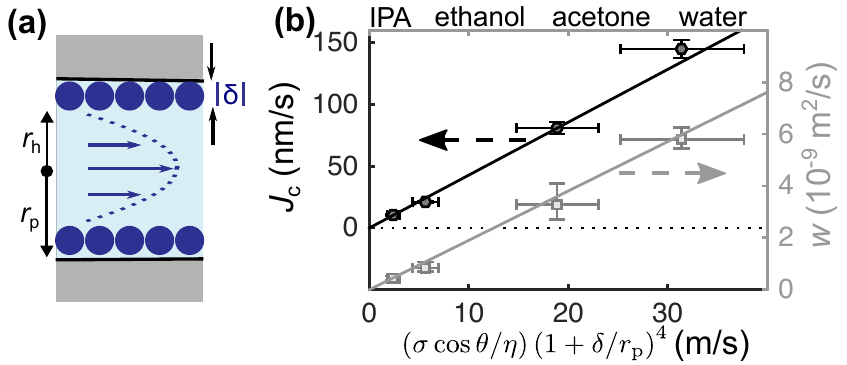}
			\caption{\small (a) The hydraulic radius $r_\mathrm{h}$ across which the parabolic flow is established differs from the geometrical pore radius $r_\mathrm{p}$ by a slip length, $\delta$ (negative for an immobile layer). (b) Plateau-drying (black circles) -- as in Fig. \ref{fig:Drying}c, regime \ding{173} -- and imbibition (grey squares) -- as in Fig. \ref{fig:Imbibition} -- dynamics as a function of the rescaled visco-capillary driving force. Continuous black and grey lines are predictions from Eqs. (\ref{eq:FluxPlateau}) and (\ref{eq:Washburn}) respectively, using Eq. (\ref{eq:Carman}) with a unique value of tortuosity, $\tau=4.5$.}
\label{fig:Liquids}
\end{center}
\end{figure}



Eq. (\ref{eq:Carman}) is based on the assumption of a developed Poiseuille flow in the pores with a zero-velocity (no-slip) condition at the pore wall. However, large deviations to the no-slip boundary conditions have been reported at the nanoscale \cite{Bocquet2010,Huber2015} and described with the concept of slip length $\delta$ which accounts for slippage ($\delta>0$) or sticking ($\delta<0$) of the fluid at the wall. The hydrodynamic radius across which the Poiseuille flow profile is established is then $r_\mathrm{h}=r_\mathrm{p}+\delta$ and the Carman-Kozeny equation (\ref{eq:Carman}) can still be used by defining \cite{Gruener2009}
\begin{equation}
r_\mathrm{eff} = \frac{r_\mathrm{h}^2}{r_\mathrm{p}} = r_\mathrm{p} \left( 1 + \frac{\delta}{r_\mathrm{p}} \right)^2.
\label{eq:EffectiveRadius}
\end{equation}
The value $r_\mathrm{eff}=1.1$ nm estimated previously is compatible with a pore size of $r_\mathrm{p}=1.7$ nm if a negative slip length of $\delta = r_\mathrm{p} - r_\mathrm{h}=-0.3$ nm is used. Interestingly, this corresponds to the typical thickness of a monolayer of water molecules, as can be estimated from the molar volume: $d_\mathrm{w}=(v_\mathrm{m}/\mathcal{N}_\mathrm{a})^{1/3}=0.31$ nm, where $\mathcal{N}_\mathrm{a}$ is Avogadro's number. This correspondence suggests that a monolayer of water molecules is immobilized at the pore wall (Fig. \ref{fig:Liquids}a), as has been proposed for nanoporous glasses \cite{Gruener2009,Xu2009}.

To investigate this hypothesis further and isolate the contributions of boundary conditions from those of the geometrical tortuosity, we measured both $J_\mathrm{c}$ (plateau-drying) and $w$ (imbibition) for 3 additional liquids (acetone, ethanol, and isopropanol) of larger size ($d=0.46\--0.50$ nm) and a variety of surface tensions and viscosities (Table S1 in \cite{Note1}). From Eqs. (\ref{eq:Carman}--\ref{eq:EffectiveRadius}), both rates should vary in proportion to the following grouping of liquid-dependent properties: $(\sigma \cos \theta / \eta) \times (1+\delta/r_\mathrm{p})^4$ with $\delta=-d$.  The plot in Fig. \ref{fig:Liquids}b shows that the rates collapse onto straight lines across the series of four liquids when the macroscopic values of $\sigma$, $\theta$, and $\eta$ and $\delta=-(v_\mathrm{m}/\mathcal{N}_\mathrm{a})^{1/3}$ are used, and with $r_\mathrm{p}=r_\mathrm{p,c}=1.7\pm0.2$ nm. This collapse, independent on the choice of $\tau$, confirms that sticking of a mono-molecular layer can explain the observed dynamics.  From the best-fit slope of each line, we can also extract a consistent value of $\tau=4.5$, a physically reasonable value for isotropic pore architectures \cite{Gruener2009}. 

We note that other scenarios (see section V in \cite{Note1}) could be invoked to accommodate these observations, but all those that we have identified involve fine tuning of parameters, for example with large, liquid-dependent tortuosities, or adjusted contact angles or viscosities for each liquid. We also note that the discussion of our results, based on the consideration of a single pore size, can be extended in a completely coherent manner to a distribution of pore sizes \cite{Note1}).

We conclude that visco-capillary flows driven by large tensile stresses in pores that are only $3$ nm in diameter follow quantitatively the macroscopic laws of fluid mechanics and thermodynamics with bulk fluid parameters provided we introduce a single correction in the form of a monomolecular, immobile layer at the pore walls. This coherency, across multiple measurements and liquids, contrasts with the large uncertainties and deviations from theory reported for nanoscale flows even in much larger conduits \cite{Bocquet2010,Chauvet2012}. Our use of steady state drying allowed us to verify that the thermodynamic stress implied by Kelvin equation expresses itself as a mechanical pressure in the liquid even in pores that measure just $6\--10$ molecules in diameter. While this mechanical equivalence of the Kelvin stress has been shown to hold in the static deformation of porous media submitted to drying \cite{Amberg1952}, our results provide the first demonstration in a dynamic context (permeation flows) and across an unprecedented range of stresses (down to  $\simeq - 100$ MPa). In technological contexts, the two regimes of passive evaporation-driven flow elucidated here have direct implications for the precise control of fluid flow in micro and nanosystems \cite{Bocquet2010} and membrane science \cite{Fornasiero2008}. In geophysical contexts, our experimental platform provides a new basis for evaluating the role of capillary phenomena in controlling the mobility of fluids within reservoirs exploited in, for example, hydraulic fracturing \cite{Birdsell2015} and sequestration of carbon dioxide \cite{Saadatpoor2010}.

\section*{Acknowledgements}

The authors thank Eugene Choi for the measurement of nitrogen isotherms and Glenn Swan for technical assistance. This work was supported by the National Science Foundation (IIP-1500261), the Air Force Office of Scientific Research (FA9550-15-1-0052), The U.S. Department of Agriculture (2015-67021-22844) and the Camille Dreyfus Teacher-Scholar Awards program and was performed in part at the Cornell NanoScale Facility, a member of the National Nanotechnology Infrastructure Network (National Science Foundation; Grant No. ECCS-15420819).


\bibliography{NanoFluidicsReferences_v12c}

\end{document}